# A method for assessing the spatiotemporal resolution of Structured Illumination Microscopy (SIM)

ABDERRAHIM BOUALAM[1] AND CHRISTOPHER J ROWLANDS[1,*]

[1]*Department of Bioengineering, Imperial College London, South Kensington, London, SW7 2BP, UK*
**c.rowlands@imperial.ac.uk*

**Abstract:** A method is proposed for assessing the temporal resolution of Structured Illumination Microscopy (SIM), by tracking the amplitude of different spatial frequency components over time, and comparing them to a temporally-oscillating ground-truth. This method is used to gain insight into the performance limits of SIM, along with alternative reconstruction techniques (termed 'rolling SIM') that claim to improve temporal resolution. Results show that the temporal resolution of SIM varies considerably between low and high spatial frequencies, and that, despite being used in several high profile papers and commercial microscope software, rolling SIM provides no increase in temporal resolution over conventional SIM.

## 1. Introduction

In wide-field microscopy, the sample is illuminated with uniform intensity, and the spatial resolution of the observed fluorescence distribution is limited by the microscope's pass band – the microscope functions as a low-pass filter, attenuating higher spatial frequencies until they cannot be observed and cannot contribute to the image (see the observable region in Fig. S1). This inability to resolve high-spatial-frequency components is known as the diffraction limit. Super-resolution structured illumination microscopy (SIM) is a widely-used super-resolution optical microscopy technique which enables up to two-fold improvement in lateral resolution (1), over and above this diffraction limit. Instead of illuminating the sample uniformly, a series of interference patterns at different phases and angles is used (1, 2), as shown in Fig. 1. This high-spatial-frequency illumination mixes with the spatial frequency content of the fluorophores in the sample, producing sum and difference frequencies which are then low-pass filtered by the microscope's pass band. High spatial frequencies which were previously beyond the diffraction limit can then be inferred. Since at least three different angles and phases are needed to adequately cover the frequency domain, the reconstruction of a super-resolved image typically requires a data-set of at least nine frames (3, 4). The reconstruction is achieved computationally, using a suitable algorithm (5-7).

During SIM acquisition, the fluorescence emission distribution consists of the sample fluorophore distribution multiplied by the illumination intensity pattern, which in the Fourier domain consists of the convolution of the Fourier transform of the sample fluorophore distribution with three spots (one central spot and two outer spots located symmetrically about the center). It is these outermost spots which 'alias' the high-frequency information from beyond the edge of the microscope's pass band back inside it, allowing its recovery.

Conventional SIM reconstruction is carried out using the successive frames independently, which means that the first SIM image is reconstructed using the first nine frames (1 to 9), the second one using the next nine frames (10 to 18) and so on (see Fig. 1, left). One might therefore intuitively expect that, since SIM requires nine frames to reconstruct one super-resolved image, its temporal resolution should be nine times worse than conventional microscopy. It is for this reason that techniques such as 'rolling SIM' reconstruction have been proposed, in which SIM frames are 're-used' in subsequent reconstructions in an attempt to increase the temporal resolution. Using these methods, instead of using frames 10 to 18 to create the next super-resolved image, frames 4-12 or even frames 2-10 are used, creating more super-resolved images using the same data, theoretically increasing the temporal resolution (see Fig. 1, right). This approach has been either proposed or used in several high-profile papers (8-10), and has even been included in commercial software, such as the BURST Mode employed in the software package used on the Zeiss Elyra 7 super-resolution microscope. Claims aside, it is important to assess the actual temporal resolution provided by this reconstruction method in order to determine whether it really has the benefits stated.

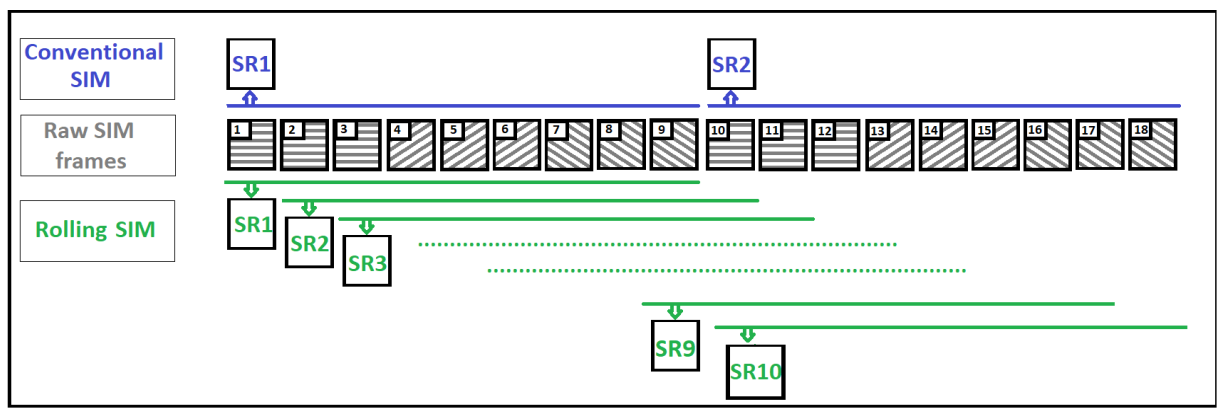

Fig. 1: Principle of conventional (top) and rolling SIM (bottom) reconstructions. In conventional SIM reconstruction, the reconstruction of the second super-resolution (SR) image uses frames 10-18 whereas rolling SIM reconstruction uses frames 2-10. With the same data set, rolling SIM can reconstruct eight additional SR images using frames 2-10, 3-11, 4-12 etc. up until 10-18.

There has been a consistent interest in the temporal resolution of SIM almost since its inception. In the original SIM implementation of Gustafsson, the striped illumination patterns were generated by moving and rotating a physical grating (3). However, the imaging speed of this technique was limited to around 1.4 s per reconstructed frame (11). Several attempts have been made to boost the imaging speed of SIM since then (11-14), notably the implementation of Heintzmann (11), which uses a spatial light modulator (SLM) instead of a physical grating. SLMs are devices which can create arbitrary, computer-controlled illumination patterns by modifying the phase or intensity distribution of the incident light field. The necessary rotations and translations of the illumination pattern can be achieved faster by reprogramming the SLM than by physically rotating or translating a material grating, resulting in this case in a super-resolved frame rate of 7.6 fps per reconstructed 2D slice. Nevertheless, the fastest implementation of SIM to date has been achieved by Curd *et al.* (14, 15) in which the reconstruction of the high-resolution SIM images is performed optically instead of using a reconstruction algorithm. This technique can achieve 3D imaging at frame rates that exceed 100fps, which corresponds to a temporal resolution below 10ms. Other attempts have been made to increase the speed of SIM without the need to improve the hardware optical design. Instead, they focus on using new reconstruction algorithms such as the rolling SIM approach mentioned earlier, or by reducing the number of frames required for the reconstruction (16-19). Notably, Ströhl and Kaminski, building on work from Ingaramo *et al.*, performed SIM reconstruction using a joint Richardson–Lucy (jRL) deconvolution algorithm (16, 20, 21) which only requires the acquisition of 3 frames (at 3 different stripes orientations, without any phase shift), allowing the temporal resolution of traditional SIM to be doubled (16).

In this paper, the temporal resolution of low and high spatial frequency information in SIM will be assessed, by exploiting the linearity property of the Fourier Transform: any uniform variation in the intensity profile in the real domain should result in the exact same variation in the amplitude/modulus of the whole Fourier spectrum (i.e. this variation should be the same for every spatial frequency). The experiment is performed *in silico* by modulating the intensity of each frame with a known sine wave and taking the normalized root-mean-squared-error between this known sine wave and the temporal variation of each spatial frequency component in the reconstructed super-resolution images. By taking this error over a number of oscillation periods, it is possible to assess the temporal resolution for different spatial frequency components independently (see Fig. 2).

## 2. Simulation details

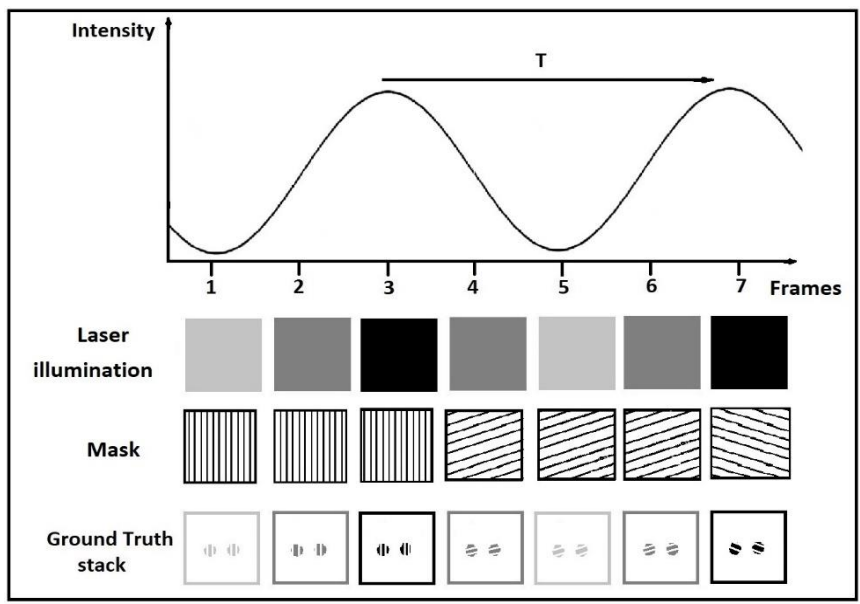

Fig. 2: Illustration of how the SIM frames are generated. The simulated laser illumination varies sinusoidally over time, and so therefore does the average intensity of the ground-truth. The resulting SIM frames are then generated by multiplying the ground-truth frames with the intensity-modulated masks.

A microscopy sample with high spatial frequency support was simulated by randomly assigning 1% of pixels in a blank image to an intensity of 1 (in the case of simulations with no noise) or the stated maximum signal intensity (in the case of simulations with noise). The intensity of this image was modulated over time by an offset sine wave $G_{T_m} = 0.75 + \frac{\sin\left(\frac{2\pi}{T_m}t\right)}{4}$ where $t$ is the time, $T_m$ is the modulation period, and $G_{T_m}$ is the ground-truth temporal modulation. $G_{T_m}$ oscillates between 0.5 and 1, in order to avoid artefacts caused by frames with zero intensity were it to oscillate between 0 and 1. An illustration of this process can be seen in Fig. 2. Raw SIM frames were created from this stack by multiplying the frames with the required intensity-modulated sinusoidal illumination profile, followed by simulation of a microscope imaging process by convolution with a simulated detection point-spread function (PSF). The simulated PSF was obtained by taking the inverse Fourier Transform of the 2D autocorrelation of a circular aperture, which corresponds to the simulated optical transfer function (OTF) of a microscope (22). At this point, Poisson noise was added if required, by drawing samples from a Poisson distribution for each pixel, with the expected value of the Poisson distribution equal to the pixel value. Finally, high-resolution (HR) SIM images were created following the interleaving reconstruction approach, i.e: the first HR image was reconstructed using frames 1 to 9, the second using frames 2-10 etc. (note that every ninth frame of this sequence corresponds to a conventional SIM frame). Simulations were repeated with 512 different temporal modulation periods, uniformly distributed between $T_m=1$ and $T_m=512$ frames. Image stack simulations and reconstructions were performed using SIMply, a SIM toolbox written in MATLAB. SIMply is an implementation of the original Gustafsson reconstruction method (3), so results are applicable to many, if not all other SIM reconstruction codes.

### 3. Method for deriving the spatiotemporal resolution of SIM

The spatial variation of light intensity in an image sequence $H(x, y, t)$ can be decomposed into a finite spectrum of spatial frequencies $(f_x, f_y)$ by using the Fourier transform, as shown in equation 1. In the 2D Fourier domain, each pixel represents a unique spatial frequency with a given orientation and phase, and the modulus of the intensity $|I(f_x, f_y)|$ is proportional to the contribution of this frequency to the spectrum. The further the pixel is from the centre of the spectrum, the greater the modulus of the spatial frequency it represents. As stated previously, because of the linearity property of the Fourier transform, any uniform variation in the intensity of the image affects the amplitude of the whole Fourier spectrum in the same way (i.e. the intensities of low and high spatial frequencies should experience the same variation). By plotting the magnitude of each spatial frequency and comparing to a 'ground-truth' (consisting of a plot of the laser intensity over time) it is possible to assess the temporal resolution, i.e. the ability for the reconstructed image to replicate different temporal frequencies resulting from the externally-modulated signal.

(1) $$I(f_x, f_y, t) = FFT_{2D}(H(x, y, t)) = \sum_{x=0}^{N-1} \sum_{y=0}^{N-1} H(x, y, t) e^{\frac{-2\pi i \cdot (f_x \cdot x + f_y \cdot y)}{N}}$$

where $N$ is the number of pixels in each direction. For each modulation period $T_m$, we compute the error in the Fourier domain, by taking the difference between the normalized Fourier spectrum magnitude $\widetilde{I_{T_m}(f_x, f_y}, t) = \frac{I_{T_m} - \overline{I_{T_m}}}{\sigma_{I_{T_m}}}$ and the ground-truth $G_{T_m}$ (see equation 2); $\overline{I_{T_m}}$ and $\sigma_{I_{T_m}}$ are the mean and the standard deviation respectively of $I_{T_m}(f_x, f_y, t)$.

(2) $$Error_{T_m}(f_x, f_y, t) = \widetilde{I_{T_m}(f_x, f_y}, t) - G_{T_m}(t)$$

Afterwards, we compute the root-mean square error $RMSE_{T_m}(f_x, f_y)$ for each $f_x$ and $f_y$ over all $t$:

(3) $$RMSE_{T_m}(f_x, f_y) = \sqrt{\frac{\sum_{t=0}^{t=t_{max}} Error_{T_m}(f_x, f_y, t)^2}{n_t}}$$

where $n_t$ is the number of time steps. We repeat this process for every modulation period $T_m$. Finally, in order to be able to visualize the RMSE as a function of both the spatial frequency magnitude $f_s = \sqrt{f_x^2 + f_y^2}$ and the modulation periods, we compute the radial average of the RMSE for each modulation period $T_m$:

(4) $$\overline{RMSE}(T_m, f_s) = \frac{1}{card(C_r)} \sum_{P \in C_r} RMSE_{T_m}(P)$$

with $C_r$ being the set of pixels located within the range r to r+1 from the centre of the image and $card(C_r)$ being the cardinality, or number of elements of this set (i.e. number of pixels in the range r to r+1).

It should be highlighted that, in some cases, the RMSE might be considered an overly-strict measure of the ability of a system to resolve a particular temporal frequency. Specifically, it is sensitive to lags between $I_{T_m}(\widetilde{f_x, f_y}, t)$ and $G_{T_m}(t)$ which, in some cases, may not be important. These lags can lead to an apparent increase the value of the RMSE, even though the temporal profile might appear very similar to the ground-truth (e.g. Fig. 3, bottom-left). To suppress this dependency of the RMSE on the time-lag, we also used an alternative metric, in which the error is defined not as the difference between $I_{T_m}(\widetilde{f_x, f_y}, t)$ and $G_{T_m}(t)$, but rather as the dot product $\dot{X}$ between the discrete Fourier transforms in time of these spectra. This permits quantification of their similarities without the effect of phase:

$$(5) \quad \dot{X} = \sum_{t=0}^{N-1} \left| FFT_{1D}\left(I_{T_m}(\widetilde{f_{x'}, f_{y'}}, t)\right) \right|^2 \cdot \left| FFT_{1D}(G_{T_m}(t)) \right|^2$$

## 4. Results and discussion

### 4.1 Comparison of the temporal resolution of two spatial frequencies

To illustrate the method for assessing temporal resolution as a function of spatial frequency, two representative low and high spatial frequencies can be plotted, to show how they can achieve different temporal resolutions. For each modulation period, the 2D Fourier Transform of each reconstructed super-resolved image was computed; the normalized temporal intensity profile of an arbitrarily randomly-selected high and low spatial frequency component (see yellow and red crosses in Fig. 3, bottom-right) were plotted against the intensity profile of the ground-truth.

As stated previously, to quantify the differences between those profiles, the normalized root-mean square error (RMSE) between the intensity profile of low and high spatial frequencies and that of the ground-truth were computed. For these isolated examples several things can be observed. Firstly, for very long periods the low frequency pixel seems to track the ground truth accurately, but the high-spatial-frequency appears to have a noticeable 'phase lag' – the functional form of the intensity modulation appears correct, but with a small time delay relative to the ground truth (and the low-spatial-frequency component). Conversely, for a modulation with a period of 9 frames, both low-spatial-frequency and high-spatial-frequency results are essentially random. This is consistent with the fact that SIM reconstruction acts as a temporal low-pass filter, and in this case the period of the oscillation is equal to the filter window and therefore cannot be resolved.

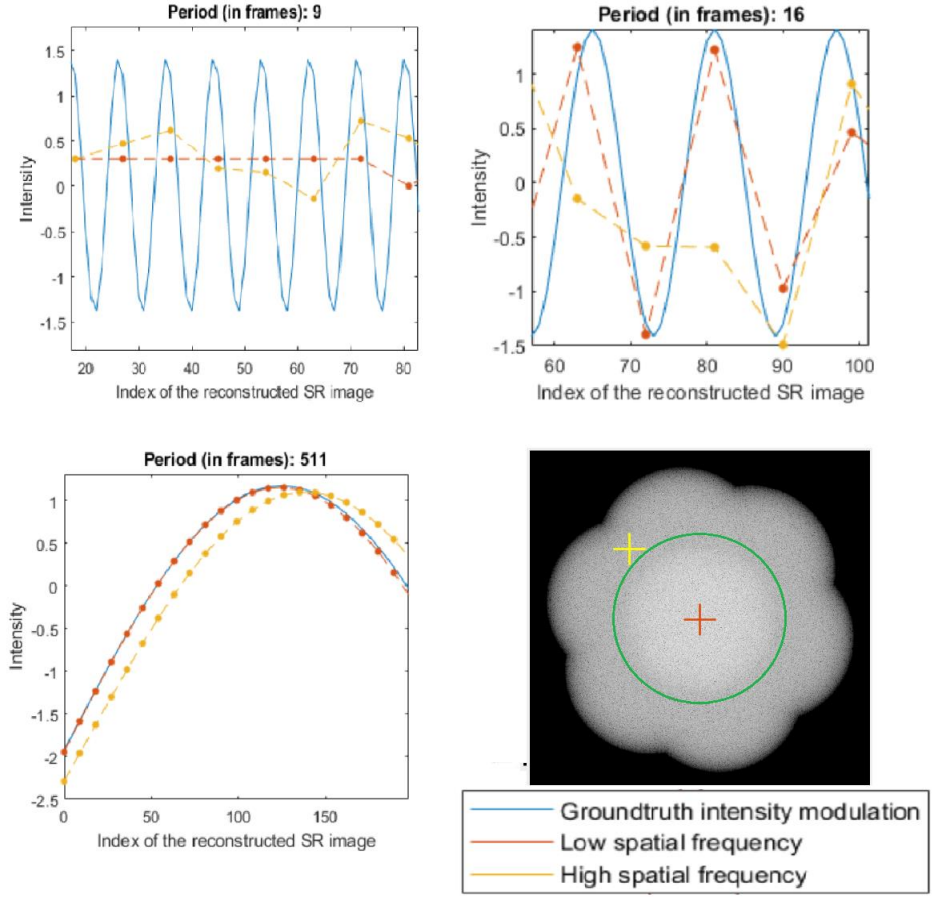

Fig. 3: Plotting the intensity of a high-frequency versus a low-frequency component for example modulation periods of 9, 16 and 511 frames. Inset bottom right is an illustration of the Fourier spectrum of a reconstructed SIM image, with the locations of the low (red cross) and high (yellow cross) spatial frequencies which are tracked. The green ring demarcates the widefield diffraction limit.

The results for a 16 frame temporal period are the most telling; they suggest that the temporal resolution of high spatial frequencies is worse than for low spatial frequencies (a visual representation of this data can be seen in Fig. S2). Exploring this phenomenon requires assessing many spatial frequencies simultaneously.

### 4.2 Estimating temporal resolution as a function of spatial frequency magnitude

To assess all spatial frequencies simultaneously, for each modulation period, an image stack is produced containing the magnitude of the FFT for each spatial frequency. For each frame, the root-mean-squared error between the normalized ground truth and the intensity over time of each spatial frequency is taken. Spatial frequencies with the same modulus are then averaged together (see Fig. S3); the mean RMSE can be plotted as a function of the radius (in pixels) from the origin, and the period of the modulation (see Fig. 4 left).

Fig. 4 highlights a number of trends. Firstly, it is clear that as the period of the modulation increases, the RMSE drops for all spatial frequencies; the temporal resolution is therefore improved. This is largely expected, but even for very large periods, the high-frequency components have a larger error than the low-frequency components. This is likely due to the fact that these frequencies are, on average, sampled less regularly than lower spatial frequencies; low spatial frequencies appear in every frame, but super-resolved frequencies by definition lie outside the OTF boundary and are only sampled when the required SIM pattern aliases them into the pass-band of the microscope. This reflects another feature of the data, which is that for all periods, the RMSE is worse for higher spatial frequencies, with a transition approximately where the spatial frequency magnitude corresponds to the period of the illuminating SIM fringes.

As noted previously, the use of normalized RMSE as a metric for assessing temporal resolution could be overly-strict in some cases, and an alternative metric which just measures whether a temporal frequency modulus can be resolved (but not necessarily with the correct phase) may be acceptable for certain niche applications. Nevertheless, the results (see Fig. 4, right) show that, even under this more generous condition, the conclusions are broadly similar. The temporal resolution at low temporal modulation frequencies is preserved for both low and high spatial frequencies, but as the temporal frequency increases, high spatial frequencies degrade more rapidly than low spatial frequencies. This demonstrates that any 'phase lag' of the reconstructed temporal profiles explains only a small part (if any) of the observed difference between the temporal resolution of high and low spatial frequencies.

### 4.2 Effect of noise

While it is enlightening to examine noise-free SIM reconstruction, in reality all experiments have noise, and it is important to consider whether this affects the results obtained above. For this analysis, a purely Poisson noise model is used; this represents a best-case scenario, in which camera noise and any excess amplification noise is negligible. The noisy data is simulated by setting the intensity of the isolated point-sources to a specified value of 'Max signal' *before* applying the SIM illumination pattern and blurring using the simulated microscope OTF; the noise is then incorporated by drawing values from a Poisson distribution for each pixel in the resulting blurred image. The data are then reconstructed in an identical manner to before.

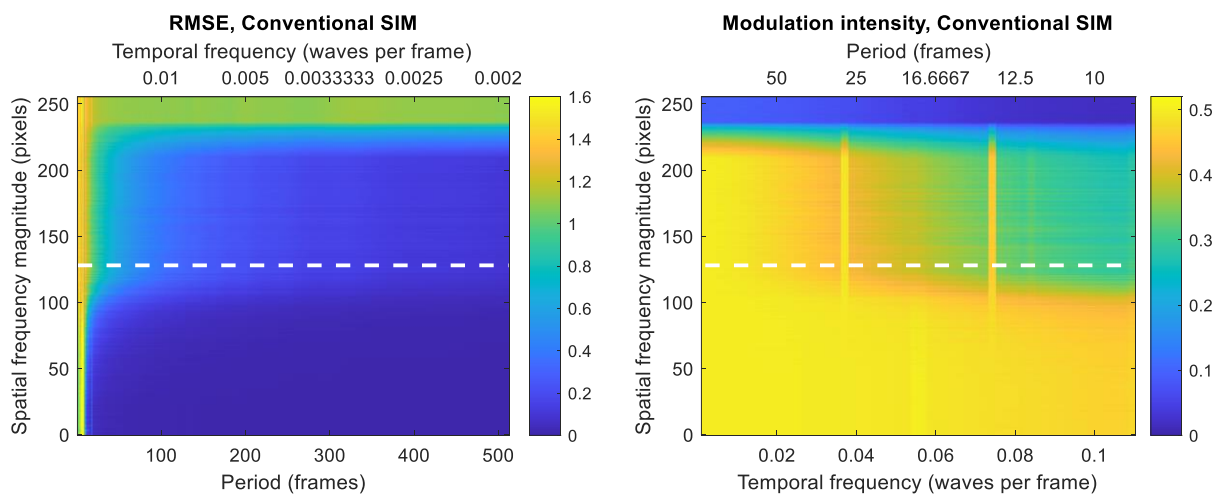

Fig. 4: Left: RMSE between the average intensity of different Fourier spatial frequencies (y axis) of the reconstructed image and the temporal intensity modulation of the raw SIM frames for different modulation periods (x axis). The intensity of different spatial frequency components is averaged along a certain radius in the Fourier domain (y-axis). The bigger the radius, the higher the spatial frequency. The horizontal dashed line shows the theoretical diffraction limit. Right: Normalized temporal frequency magnitude at the temporal oscillation frequency, as a function of temporal frequency and spatial frequency magnitude. Note that the X axis is in monotonically-increasing units of temporal frequency, because the temporal periods are constrained to an even integer divisor of 512 (the number of simulated timepoints).

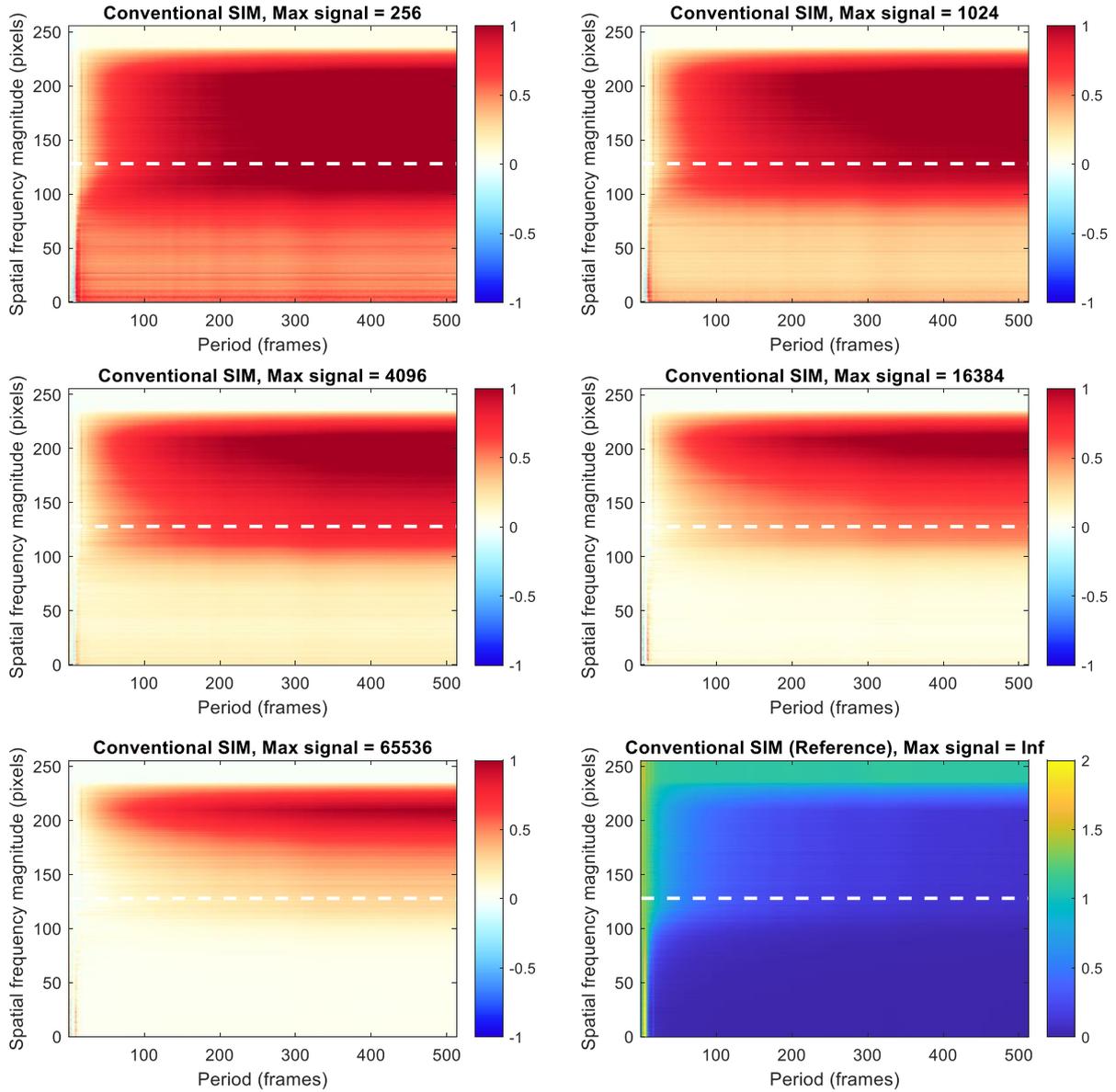

Fig. 5: Noise sensitivity of SIM reconstruction, as a function of spatial frequency magnitude and temporal oscillation period. All data presented as RMSE relative to the noise-free reference image, bottom right. Note the different scale on the color bars for the reference data relative to the noise data.

The results (seen in Fig. 5) show that the predominant effect of noise is to degrade the performance of the SIM reconstruction at longer periods; performance at shorter periods is already quite poor, so the additional noise does not degrade the reconstruction to the same degree. More notable however is a clear difference between the low ('widefield') spatial frequencies and the high ('super-resolved') spatial frequencies. Low spatial frequency performance is degraded by a broadly constant value, regardless of the period of the temporal oscillation, but the high spatial frequencies are reconstructed proportionally poorly at longer temporal oscillation periods, compared to the reference (zero noise) reconstruction. Notably, this phenomenon doesn't affect all 'super-resolved' spatial frequencies equally (unlike the 'widefield' spatial frequencies) – higher 'super-resolved' spatial frequencies are degraded proportionately more.

This has implications for SIM imaging of dynamic events. Good super-resolution imaging normally requires bright images, but as the temporal frequency content of the dataset increases, the benefit of a high signal-to-noise ratio is lost; simultaneous reconstruction of both high spatial and high temporal frequencies is poor regardless of signal-to-noise ratio. Counterintuitively, for moderate temporal frequencies, the photon budget of the system may mean that better temporal resolution can actually be achieved by imaging *more slowly*, but with better signal-to-noise ratio.

### 4.3 Effectiveness of Rolling SIM reconstruction

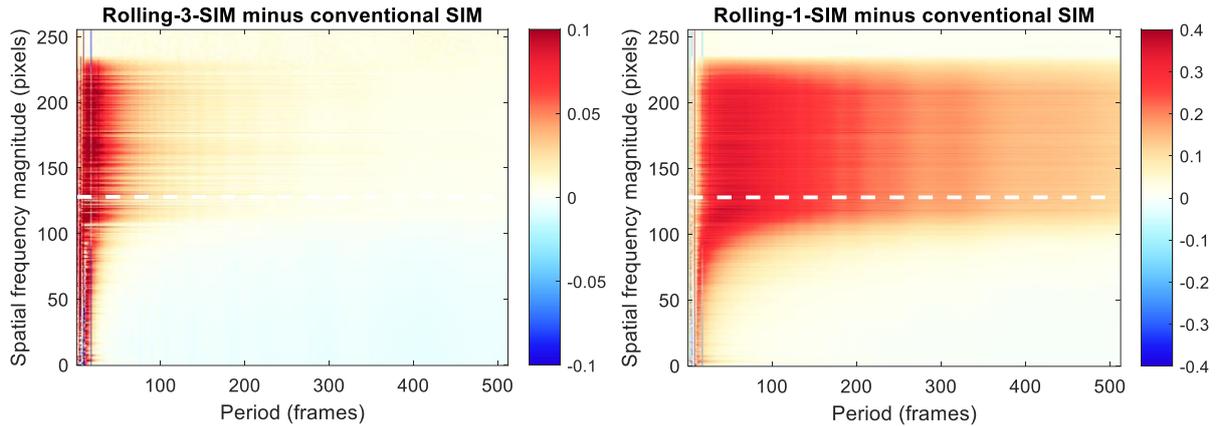

Fig. 6: Difference between the temporal resolution of Rolling-3-SIM and conventional SIM, and of Rolling-1-SIM and conventional SIM, obtained by subtracting the RMSE for conventional SIM from that of Rolling SIM.

As noted previously, one approach to increasing the temporal resolution of SIM is to perform a 'rolling' reconstruction, in which data from previous super-resolved frames are 're-used'. Whether this provides a true increase in temporal resolution is subject to debate. Importantly, it is logical to assert that, in order to be considered successful, rolling SIM should increase the temporal resolution of the *super-resolved* spatial frequency components (because if conventional widefield resolution were acceptable for a given application, there would be little point in using SIM).

Two versions of rolling SIM will be explored; in 'rolling-3-SIM', a SIM reconstruction is performed every three frames, ensuring that all frames with the same angle but different phases are taken sequentially (i.e. frames 1-9, 4-12, 7-15 and so on, but not frames 2-10). In 'rolling-1-SIM' the SIM reconstruction is performed every frame, accepting that in two thirds of cases, one set of three phases will be split between the last and first frames of the reconstructed set. One might therefore expect that the temporal resolution of rolling-1-SIM would be prone to temporal artefacts.

Fig. 6 highlights the differences between the RMSE values obtained with rolling SIM and conventional SIM. For slowly-varying temporal frequencies, the performance of rolling-3-SIM shows a subtle advantage at lower spatial frequencies, but an equally subtle disadvantage at higher spatial frequencies, with no significant performance benefit. At low temporal periods however (where improved temporal resolution is most necessary) there is a pronounced degradation in the RMSE performance, leading to the conclusion that rolling-3-SIM should not be employed in order to increase a SIM microscope's temporal resolution.

As predicted, the performance of rolling-1-SIM is clearly worse than rolling-3-SIM, most likely due to reconstruction artefacts, owing to the previously-identified split phases in two-thirds of the reconstructions. Notably, there does not even appear to be a small benefit at low spatial frequencies (in the manner of rolling-3-SIM), which means this reconstruction method is not recommended even in the rare cases when the low spatial frequencies are of interest in a super-resolution image.

## 5. Conclusion

The effective spatial resolution and information content achievable by SIM and other super-resolution techniques have been subject to intense scrutiny in the last decade (23-25). Though super-resolution methods open the door to the investigation of biological processes with unprecedented detail, they also significantly increase the technical complexity of microscopes and image reconstruction techniques. Hence, in order to acquire and interpret high-quality super-resolution data, it is essential to understand the specific limitations and distortions imposed by a given microscopy technique. Every microscope, even the simplest one, performs a spatial mapping of the sample structure onto an image, and biological interpretation relies on understanding this mapping. In this work, we addressed this mapping from a temporal perspective, on which interpreting biological significance relies just as much.

In summary, a method for assessing the spatiotemporal resolution of optical super-resolution microscopy has been developed and has been used to investigate the spatiotemporal resolution properties of Structured Illumination Microscopy. Notably, it was found that the temporal resolution of the system is not the same for both the low and high spatial frequencies; the low-frequency information (that would ordinarily be gathered by a widefield microscope) has superior (and more consistent) temporal resolution, whereas the super-resolved spatial frequencies are much more prone to temporal artefacts. The effect of noise on temporal resolution is very pronounced, and very high signal-to-noise ratios are needed to resolve temporal periods comparable with the 9-frame period of a SIM measurement. Furthermore, it was found that the interleaved reconstruction of rolling SIM does not support any increase in temporal resolution of the super-resolved information, and in fact results in a small (but largely insignificant in the case of rolling-3-SIM) reduction in temporal resolution for these spatial frequency components.

## Acknowledgements

CJR acknowledges support from the Wellcome Trust (212490/Z/18/Z), EPSRC (EP/S016538/1), BBSRC (BB/T011947/1) and the Imperial College Excellence Fund for Frontier Research. AB is grateful for a PhD studentship from the Imperial College Department of Bioengineering. Both authors are indebted to Debora Machado Andrade Schubert for proofreading and commenting on the manuscript.

# A method for assessing the spatiotemporal resolution of Structured Illumination Microscopy (SIM): supplemental document

ABDERRAHIM BOUALAM[1] AND CHRISTOPHER J ROWLANDS[1,*]

[1]*Department of Bioengineering, Imperial College London, South Kensington, London, SW7 2BP, UK*
**c.rowlands@imperial.ac.uk*

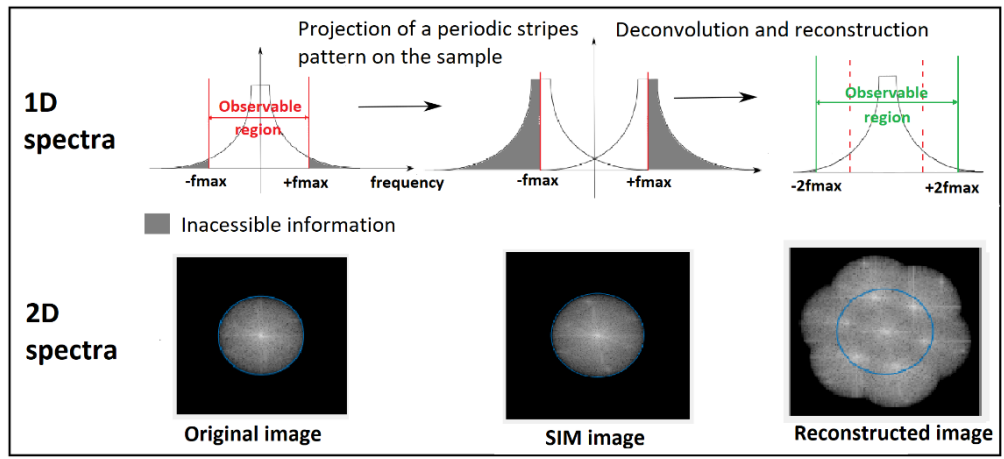

Fig. S1: SIM resolution enhancement, visualized in the Fourier domain. Left: 1D and 2D spectra of a widefield image. Middle: 1D and 2D spectra of the widefield image illuminated with a striped pattern. Right: 1D and 2D spectra of a high-resolution SIM reconstructed image. SIM imaging gives access to an observable region approximately twice as large as that of widefield image (blue circle). Modified from *(3)*.

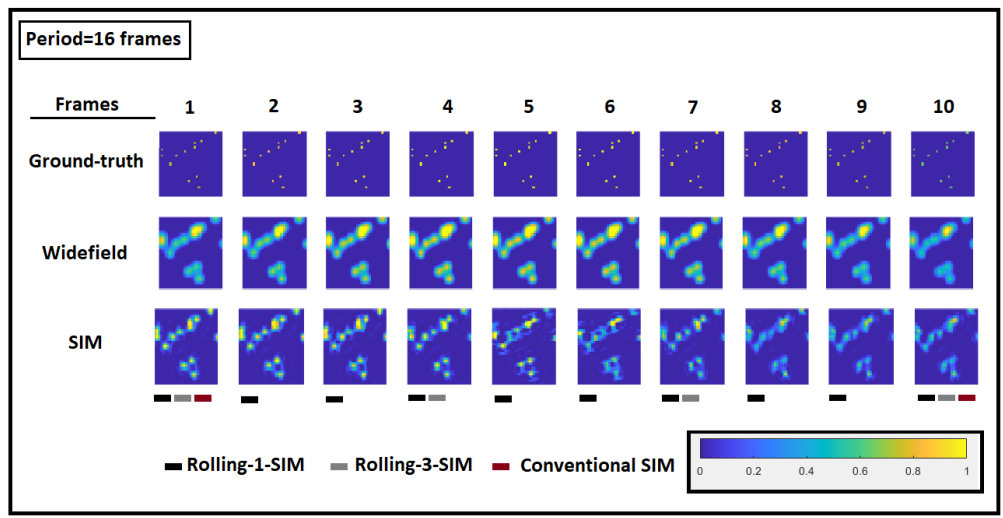

Fig. S2: Illustration of the imaging performance of SIM reconstruction in the presence of pronounced high-temporal-frequency oscillations. In an ideal case, all reconstructed frames would be identical in appearance, because the SIM reconstruction would not be affected by temporal modulation.

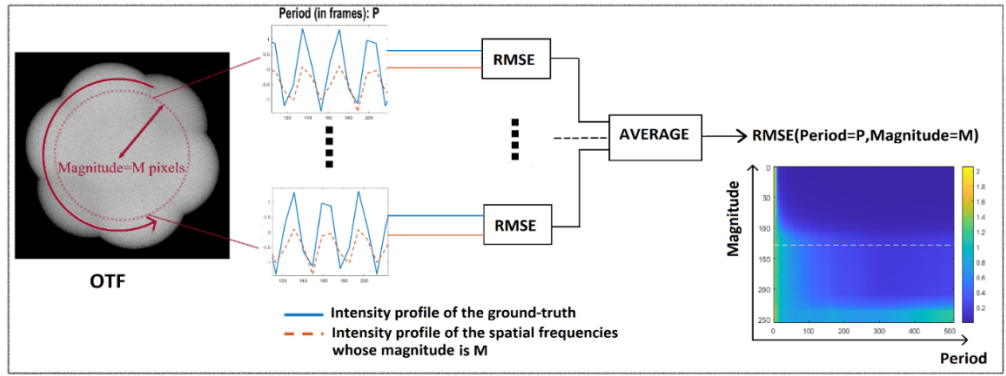

Fig. S3: Calculation of the RMSE metric. RMSE between the ground-truth intensity profile and that of spatial frequencies with the same magnitude M are averaged radially (see the red circular arrow on the OTF) and plotted for different modulation periods P (see right hand side plot)